\documentclass[twocolumn,prb,showpacs,preprintnumbers,amsmath,amssymb,letterpaper]{revtex4}
\usepackage{subfigure}
\usepackage{graphicx}% Include figure files
\usepackage{dcolumn}% Align table columns on decimal point
\usepackage{bm}% bold math
\usepackage{setspace}
\begin{document}

%\preprint{APS/123-QED}

\title{Signatures of superconducting gap inhomogeneities \\ in optical properties} % Force line breaks with \\

\author{J. P. F. LeBlanc}
\email{leblanc@physics.uoguelph.ca}
% \altaffiliation[Also at ]{Physics Department, University of Guelph.}%Lines break automatically or can be forced with \\
\author{E. J. Nicol}%
\affiliation{Department of Physics, University of Guelph,
Guelph, Ontario N1G 2W1 Canada} 
\affiliation{Guelph-Waterloo Physics Institute,
University of Guelph, Guelph, Ontario N1G 2W1 Canada}

\author{J. P. Carbotte}
\affiliation{Department of Physics and Astronomy, McMaster
University, Hamilton, Ontario N1G 2W1 Canada}
\affiliation{The Canadian
Institute for Advanced Research, Toronto, ON M5G 1Z8 Canada}

\date{\today}% It is always \today, today,

\begin{abstract}

Scanning tunneling spectroscopy applied to the high-$T_{c}$ cuprates has 
revealed significant spatial inhomogeneity on the nanoscale.  Regions on 
the order of a coherence length in size show variations of the magnitude of 
the superconducting gap of order $\pm20\%$ or more.  An important unresolved 
question is whether or not these variations are also present in the bulk, 
and how they influence superconducting properties. 
As many theories and data analyses for high-$T_{c}$ superconductivity assume spatial homogeneity of the gap magnitude, this is a pressing question. 
We consider the far-infrared optical conductivity and evaluate, within 
an effective medium approximation, what signatures of spatial variations 
in gap magnitude are present in various optical quantities. In addition
to the case of d-wave superconductivity, relevant to the
high-$T_c$ cuprates, we have also considered s-wave gap symmetry in
order to provide expected signatures of inhomogeneities for superconductors
in general.  While signatures of gap inhomogeneities can be strongly manifested in s-wave superconductors, we find that the far-infrared optical conductivity in d-wave is robust against such inhomogeneity.

\end{abstract}

\maketitle

\section{Introduction}

There is considerable evidence from scanning tunneling microscopy (STM) that 
some cuprates are intrinsically inhomogeneous on the nanoscale.  While Bi$_{2}$Sr$_{2}$CaCu$_{2}$O$_{8+\delta}$ (BSCCO) has been extensively studied, \cite{pan:2001,lang:2002,howald2:2003,fang:2006,gomes:2007,mcelroy:2005,mcelroy:science:2005,pasupathy:2008} 
spatial inhomogeneities exist as well in La$_{2-x}$Sr$_{x}$CuO$_{4}$ 
(Ref.~\cite{kato:2005}) and in electron doped systems such as 
Pr$_{0.88}$LaCe$_{0.12}$CuO$_{4}$.\cite{niestemski:2007}  On the other hand, 
the inhomogeneities seen in STM can be affected through sample preparation 
methods \cite{hoogenboom:2003}.  A critical question is whether these 
inhomogeneities exist only on the surface layer to which STM is sensitive, 
or are also present in the bulk.  Evidence that they are not present in the 
bulk is provided by NMR data on YBa$_{2}$Cu$_{3}$O$_{7-\delta}$ (YBCO) 
\cite{bobroff:2002}.  Also, Loram et al. \cite{loram:2004} have argued that, 
even in BSCCO, bulk inhomogeneities on the scale indicated in STM experiments 
are inconsistent with specific heat data; although this view has been 
challenged recently by Andersen et al. \cite{andersen:2006} who concluded that 
nanoscale inhomogeneity in BSCCO is a bulk property.  In view of this 
conflict, it is clearly important to look for other possible probes of bulk 
nanoscale variation.  Such a probe is infrared absorption.  In this paper, 
we consider the effect of nanoscale regions on optical properties with an 
aim at identifying their signatures in these quantities.

\begin{figure}
\subfigure[]{
	\includegraphics[width=0.22\textwidth]{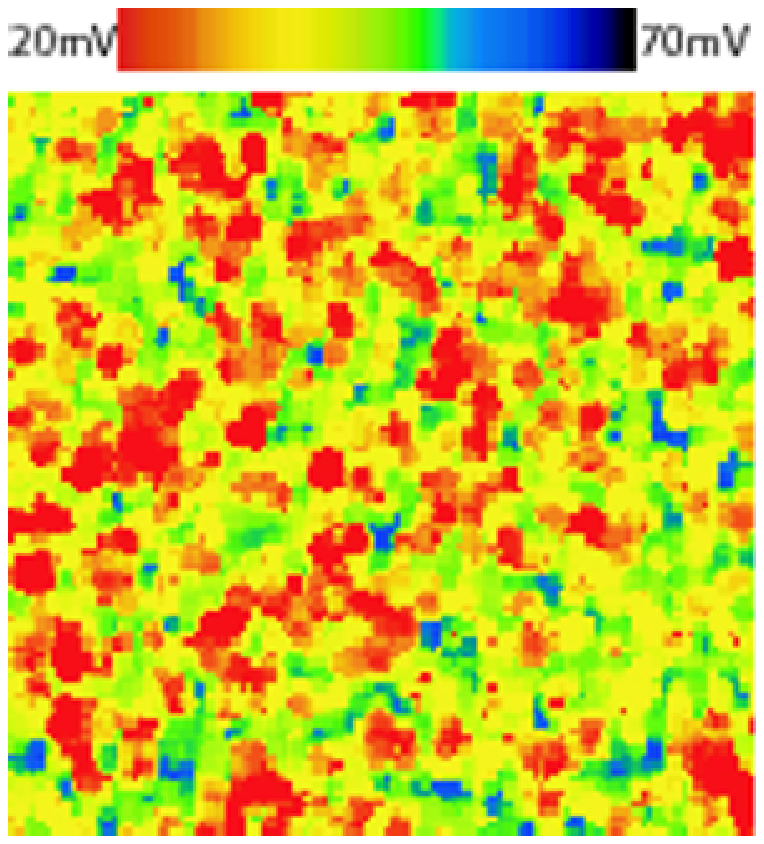}
\label{fig:gapmaps:a}
	}
\subfigure[]{
\includegraphics[width=0.22\textwidth]{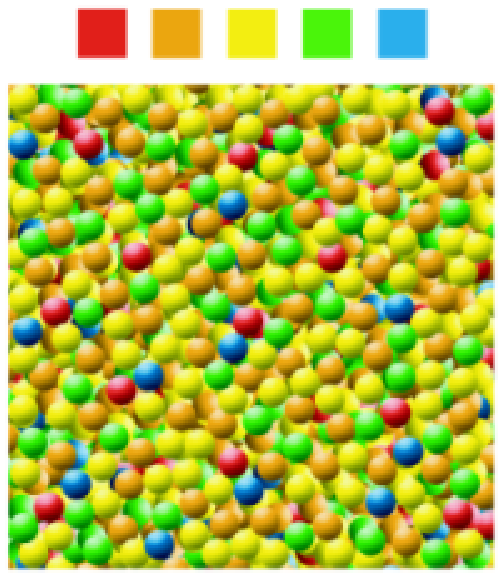}
\label{fig:gapmaps:b}
	}
	\subfigure[]{
\includegraphics[width=0.22\textwidth]{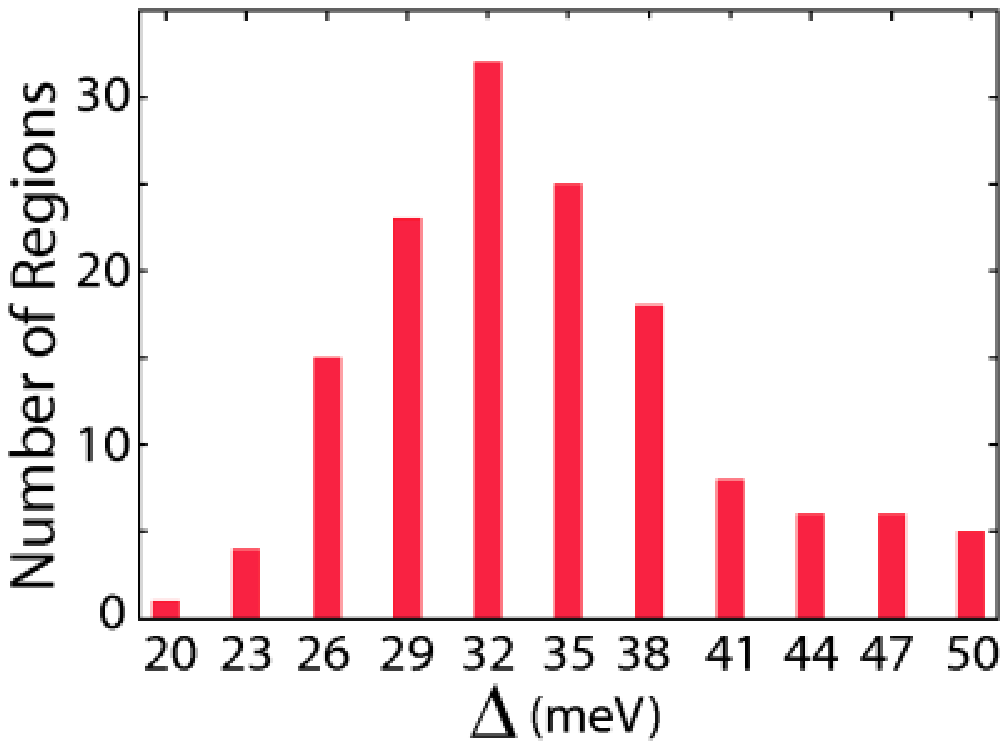}
\label{fig:gapmaps:c}
	}
	\subfigure[]{
\includegraphics[width=0.22\textwidth]{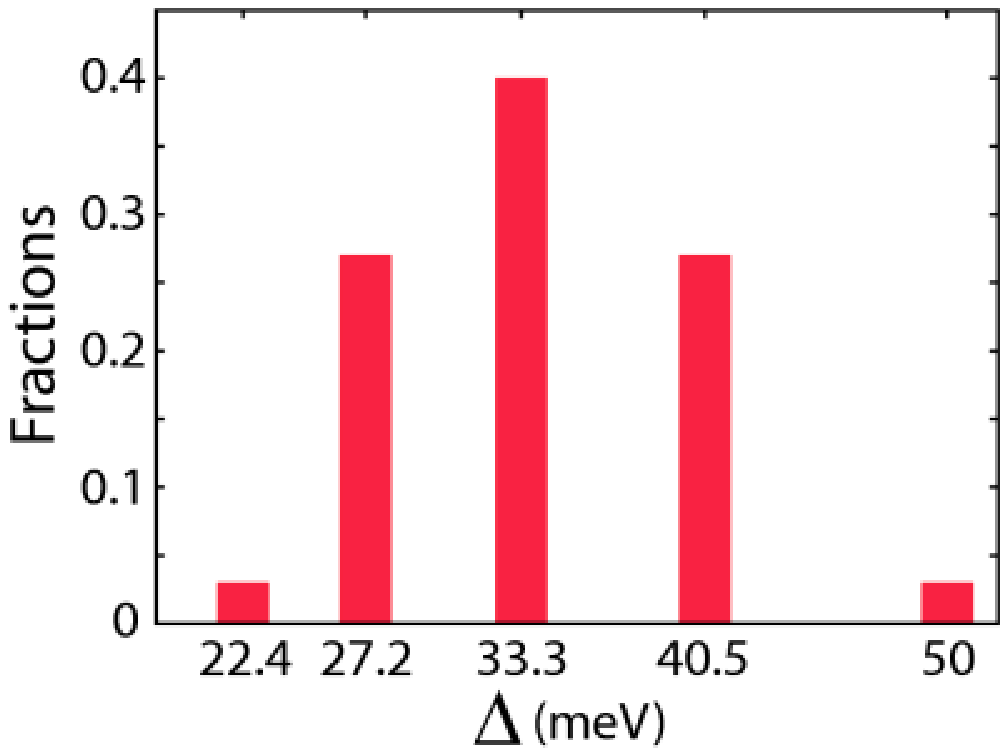}
\label{fig:gapmaps:d}
	}
\caption{\label{fig:gapmaps} (Color online) $\mathbf{(a)}$ STM gap map of surface of overdoped (OD) BSCCO with $T_{c}$=89K from Ref.~\cite{lang:2002,mcelroy:2005}.$\mathbf{(b)}$ Schematic of system comprised of spherical particles distributed at random as dictated by histogram (d). $\mathbf{(c)}$ Histogram of gap distribution shown in (a), adapted from Ref.~\cite{lang:2002}.  $\mathbf{(d)}$  Re-binning of histogram (c) to a set of five gaps.}
\end{figure}

\begin{figure}
	\includegraphics[width=0.3\textwidth]{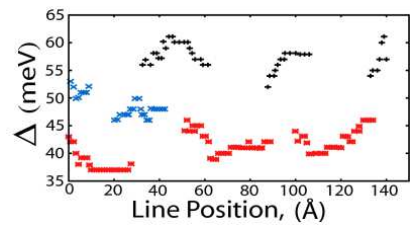}
	\caption{\label{fig:linescan} (Color online) Adapted from Ref.~\cite{lang:2002}.  Plot of superconducting gap value as a function of position along a straight line of an STM gap map.}
\end{figure}

To understand how nanoscale variations of the superconducting energy
gap can affect optical properties, we have applied 
an Effective Medium Approximation (EMA) to an inhomogenous system with a distribution 
of superconducting energy 
gaps.  The merit of this approach is that we do not commit
ourselves to any particular microscopic model of the inhomogeneity, but rather
examine the macroscopic electrodynamics with a minimum of assumptions. In order
to consider the case of the high $T_c$ superconductors, we include 
d-wave gap symmetry in our calculations and model the gap distribution
based on BSCCO STM gap 
maps of Refs.~\cite{lang:2002,mcelroy:2005}. 
As can be seen in these and other references, 
using STM to image a surface allows for mapping 
of high and low regions of superconducting gap magnitude, $\Delta$. The lattice structure of the BSCCO compounds is such 
that the surface can cleave smoothly between adjacent Bi-O planes, 
resulting in what should be a uniform surface structure.
However, despite the supposed uniform lattice structure - that is to say 
there is no evidence of restructuring on the surface after cleaving - there 
is a spatially inhomogeneous energy gap structure at the surface.  
These inhomogeneous gap 
structures are commonly displayed in the form of a gap map, as seen in 
Fig.~\ref{fig:gapmaps:a}.  In this figure, the blue patches are regions
with a gap magnitude of about 50 meV and the red have gap values of about
20 meV. The largest number of regions occur with a gap of $\sim 32$ meV
and the variation away from this value is about $\pm 20\%$ as indicated
in Fig.~\ref{fig:gapmaps:c}. 
The diameters of these regions are typically of order 3 nm,\cite{mcelroy:science:2005} but there
is no particular pattern or underlying structure, rather 
the regions appear to be randomly distributed.
It is also important to note the discontinuous 
nature of these gap maps, as can be seen in Fig.~\ref{fig:linescan}, where a scan along a line across the surface is shown.  Here we see that the gap map contains distinct regions with nearly uniform gap values 
and consequently
does not represent a smoothly varying single gap function $\Delta(x)$, but
rather suggests separate patches, each with its own gap value.
 In implementing the EMA as 
a phenomenological model, we are assuming that these 
STM images directly reflect the 
bulk gap distribution in an attempt to find signatures of this gap 
inhomogeneity.  However, in the following, our results for far-infrared optical conductivity will show that
no strong signatures were observed in mixtures of d-wave superconductors
 and indeed 
the results of the EMA can be very well 
approximated by a single spatially averaged gap value, which provides support for the past and continuing use of theories and data analyses which assume a spatially homogeneous energy gap in the bulk of high-$T_{c}$ superconductors.  It would appear from our work that even if the inhomogeneities are present in the bulk, this assumption will be robust.  On the other
hand, we find that signatures in  
mixtures of s-wave superconductors
present themselves in the form of a distribution of gap onsets 
over the range of gaps.  This distribution results in an optical conductivity 
and optical self-energy which are distinct from a single gap system, 
and that are 
highly dependent upon the range of variation in the gap as well as the actual 
distribution present.   In the following section, we begin by
presenting the theory for the EMA and its use to obtain the optical
conductivity. In section III, we will present our results for d-wave
superconductivity based on the distribution of gaps seen in STM. Section IV
will then present results for s-wave superconductivity and we summarize our
conclusions in section V.

%\begin{figure}
%	\includegraphics[width=170mm]{X:/paper/figures/figure2-1.pdf}
%	\caption{\label{fig:2} Adapted from \cite{lang:2002}.  Plot of 
%superconducting gap value as a function of position along a straight line 
%of an STM gap-map.}
%	
%\end{figure}

%\subsection{\label{sec:level2}Motivation: STM}

\section{Effective Medium Theory}

It is of interest to us to understand in which systems and on what energy 
scales we might expect to see indications of inhomogeneities in bulk optical 
properties in order to make generalizations on, or identify, inhomogeneities 
in new materials.  For this reason, we seek a purely phenomenological method 
of examining these effects, and here we will turn to Bruggeman's EMA \cite{bruggeman:1935,infrared}.  This EMA has been 
rigorously tested and previously applied to both microscopic metallic and microscopic superconducting-normal mixtures\cite{stroud:1975,stroud:1978,stroud:1979,landauer:1952,garner:1983}.  It has also been used more recently to model\cite{barabash:2003,barabash:2003:prb} the very low frequency optical response in the cuprates and to discuss, in particular, the THz data on BSSCO from Corson et al.\cite{corson:1999,corson:2000}  In certain circumstances,
 an alternative choice of EMA might be used, the
 Maxwell-Garnet EMA,\cite{infrared} however, that EMA applies when there are a dilute
set of regions embedded in a large uniform background, which is not the
case seen here in the STM of Fig.~\ref{fig:gapmaps:a}.
In the following, we wish to make a semi-classical approximation, 
treating the system 
classically as having randomly distributed regions, each a uniform superconducting material, with properties defined by non-classical BCS results.  
The implicit assumption here is the randomness of the distribution of
regions, since it has yet to be understood how and if these 
inhomogeneities are distributed throughout the bulk material.

For understanding the optical response of the entire system, we 
now summarize the idea of the EMA. 
An electromagnetic wave, of frequency $\omega$, moving through a composite 
material will undergo multiple scattering and absorption events.  In a bulk 
material these events are accounted for as an average by means of an 
effective complex dielectric function $\epsilon_{m}(\omega)$.  If the bulk 
material is a composite, it is assumed to be composed of a number 
of regions of uniform dielectric functions $\epsilon_{i}(\omega)$.  The 
scattering caused by these regions is dependent upon both the shape and size 
of the regions (grains).  These scattering events can be taken into account by allowing each grain to become polarized, producing its own additional field.  When we extend this to a macroscopic system, we add the constraint that the entire material should have a zero net polarization, i.e.\cite{landauer:1952}
\begin{equation}
\sum_{i}{f_{i}P_{i}}=0,
\end{equation}
where $f_i$ is the relative fraction of the occurance of a certain type of grain indexed by $i$ and $P_{i}$ is that constituent's polarization.  Knowing this, we can solve for the real and imaginary parts of $\epsilon_{m}(\omega)$ in the following 
equation:
\begin{equation}
\sum_{i}{f_{i}\frac{(\epsilon_{i}-\epsilon_{m})}
{g\epsilon_{i}+(1-g)\epsilon_{m}}}=0,
\label{eq:emt}
\end{equation}
where $g$ is the depolarization factor, determined by the shape of the 
constituent grain.\cite{infrared,stroud:1975}  The form of the polarization in
Eq.~(\ref{eq:emt}) can be seen as the electric dipole contribution of 
a multipole expansion.  Subsequent terms in the polarization (magnetic dipole, electric 
quadrupole, etc.) scale as functions of the grain size.  In the spherical case 
these terms scale as $(\omega R_i/c)^2=(R_i/\lambda)^2$ where $R_{i}$ 
is the radius of the individual grains and $\lambda$ is the wavelength of 
the field.  For this case, the correction to Eq.~(\ref{eq:emt}) to next order would be:
\begin{align}
\sum_{i}f_{i}\biggl[&\frac{(\epsilon_{i}-\epsilon_{m})}{g\epsilon_{i}+(1-g)
\epsilon_{m}}\nonumber\\
&+\frac{1}{30}\biggl(\frac{\omega R_{i}}{c}\biggr)^{2}
(\epsilon_{i}-\epsilon_{m})\biggr]=0 .
\label{eq:emtrad}
\end{align}
We can see certain limits where the second term can be neglected.  
In the small grain limit, we would expect the far-infrared (long wavelength 
limit) response to be completely dominated by the first term.  As this is
the region of interest for detecting signatures of the superconducting energy gap,  
the EMA as in Eq.~(\ref{eq:emt}), with g set to a value of 1/3 (spherical grains), is appropriate for our purpose.\\
\indent To implement the above EMA, we require the frequency dependent complex
conductivity $\sigma(\omega)=\sigma_1(\omega)+i\sigma_2(\omega)$ as the
dielectric function can be written as $\epsilon(\omega)
=\epsilon_\infty+i[4\pi\sigma(\omega)/\omega]$, where $\epsilon_\infty$
is the high frequency dielectric constant.
The constituent conductivities that we use for our purpose
here are calculated according to Zimmermann et al.,\cite{zimmermann:1991} 
for the case of BCS s-wave superconductors with variable impurity scattering
and according to Sch\"urrer et al.,\cite{SSC:1998} for the BCS d-wave case.  
In this paper, 
we consider only an elastic scattering rate in the optical
conductivity and have not included inelastic scattering, as might
arise due to electron-boson interactions, such as phonons and spin fluctuations.  
For a system of  $N$ different types of superconducting grains,
 the EMA of Eq.~(\ref{eq:emt}) can be expanded to the form of an $N^{th}$ 
order complex polynomial in $\epsilon_{m}(\omega)$ or by setting 
$\epsilon_{k}=1+i(4\pi\sigma_{k}/\omega)$ ($k=1,2, \ldots,N$ and $m$) as a 
complex polynomial in $\sigma_{m}$.  For example, 
in $\epsilon$ we now have:
\begin{equation}
\epsilon^N_{m}(\omega)+a_{1}\epsilon^{N-1}_{m}(\omega)+ \cdots +a_{N}\epsilon^0_{m}(\omega)=0 ,
\label{eq:polynomial}
\end{equation}
where the coefficients $a_{i}$ vary based on the choice of $N$.  For an $N=2$ 
system, one can solve directly for $\epsilon_{m}$ by substituting in its complex form,
 $\epsilon_{m}=\epsilon_{1m}+i\epsilon_{2m}$ and solving for the real and imaginary 
parts separately.  Analytically, this becomes largely unfeasible for $N>2$.  
For this reason, considerations of composites, with a large number of 
constituents, $N$, have not been thoroughly explored.  Numerical solutions, 
however, are much more feasible.  All that is required is solving 
for the set of $a_{i}$'s for each choice of N.  This turns out to
be a difficult task for $N>3$ and it is preferable to use a 
program to solve this 
algebraically.  For example, for the N=2 case:
%\begin{flushleft}
%\begin{eqnarray}
%a_{1}&=& \frac{1}{b_0}\biggl[-f_{1}\epsilon_{1}(\omega)(1 - g)+f_{1}g\epsilon_{2}(\omega)\\
%   & &-f_{2}\epsilon_{2}(\omega)(1 - g)+f_{2}g\epsilon_{1}(\omega) \biggr],\nonumber \\
%a_{2}&=& \frac{1}{b_0}\biggl[-f_{1} \epsilon_{1}(\omega) g \epsilon_{2}
%(\omega) - f_{2} \epsilon_{2}(\omega) g \epsilon_{1}(\omega)\biggr],
%\label{eq:2graincoef}
%\end{eqnarray}
%\end{flushleft}
\begin{eqnarray}
a_{1}= \frac{1}{b_0}\biggl[&-&f_{1}\epsilon_{1}(\omega)(1 - g)+f_{1}g\epsilon_{2}(\omega)\nonumber\\
&-&f_{2}\epsilon_{2}(\omega)(1 - g)+f_{2}g\epsilon_{1}(\omega) \biggr],
\end{eqnarray}
\begin{eqnarray}
a_{2}&=& \frac{1}{b_0}\biggl[-f_{1} \epsilon_{1}(\omega) g \epsilon_{2}
(\omega) - f_{2} \epsilon_{2}(\omega) g \epsilon_{1}(\omega)\biggr],
%\label{eq:2graincoef}
\end{eqnarray}
where $b_{0}=f_{1}(1 - g) + f_{2} (1 - g)$.
We can see that these coefficients depend on the dielectric function 
$\epsilon_{i}(\omega)$ (or conductivity) of each constituent, as well as their 
relative volume fractions $f_i$ and the depolarization factor, $g$.  
These are the key pieces of information required to perform such an effective 
medium calculation, and as such, will be explained in detail for each case we 
wish to examine.  Knowing the numerical coefficients, $a_{i}$ in Eq.~\ref{eq:polynomial}, reduces the EMA calculation to the problem of numerically solving an N$^{th}$ order complex polynomial.  Such a polynomial has N solutions, of which only one is generally completely physical as $\epsilon_{2}(\omega)$ must be strictly positive for each choice of $\omega$.

\section{D-wave Energy Gap Inhomogeneities}

\begin{figure}
\includegraphics[width=0.3\textwidth]{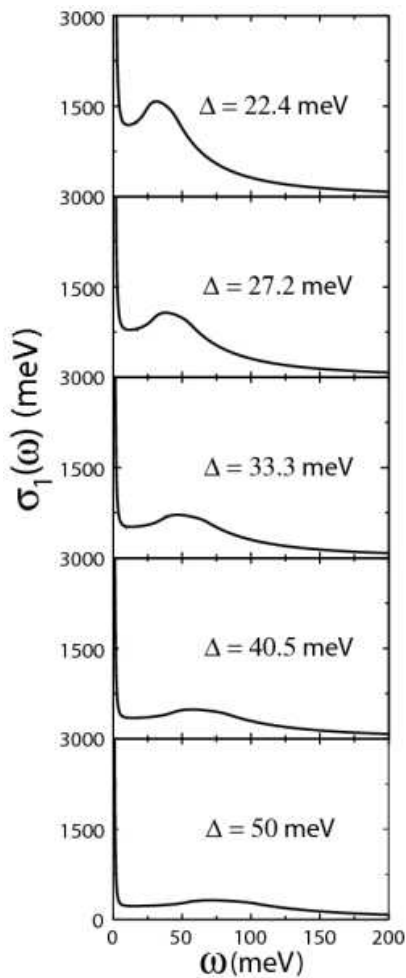}
\caption{\label{fig:5d} The real part of the optical conductivity for
a d-wave superconductor at $T=20$K for different gap values chosen
 to match those in the re-binned gap map
histogram [Fig.~\ref{fig:gapmaps:d}].  The curves shown here have $\omega_{p}=2$~eV and
 $t^{+}=0.2$~meV.}
\end{figure}

In order to determine possible signatures in the bulk due to the 
inhomogeneities seen by STM,
we created a mixture of superconducting patches which is
motivated by Fig.~\ref{fig:gapmaps:a} and shown schematically in
Fig.~\ref{fig:gapmaps:b}. We tried to maintain the same gap distribution
seen in experiment and
shown in the histogram of Fig.~\ref{fig:gapmaps:c}, but have re-binned
the distribution so that it gives only five fractions $f_i$ in order to
reduce the complexity of the numerical calcuation.  This is shown in Fig.~\ref{fig:gapmaps:d}.  Furthermore,
we assumed that each constituent has d-wave gap symmetry 
similar to what one would expect in the bulk BSCCO 
conductivity\cite{hwang:2007} and have used the BCS d-wave conductivity 
program of Ref.~\cite{SSC:1998} to provide the input conductivities to
the EMA. The BCS conductivity, while missing the inelastic component, does
exhibit similar qualitative features as
seen in experimental 
results, and can be applied with a minimal number of parameters: the energy
gap, 
$\Delta$; the plasma frequency, $\omega_{p}$; the temperature, $T$;
 and the scattering 
parameter $t^{+}=1/(2\pi\tau)$,
 where $1/\tau$ is the elastic scattering rate. 
Since we are interested in the effects of varying only the gap parameter
and we wish to limit the parameter space, we 
compute the EMA using constituents with 
similar plasma frequencies and scattering 
rates, but a range of $\Delta$ values.  
The real part of the input conductivities of the constituent superconducting
grains are shown 
in Fig.~\ref{fig:5d}.  These curves were calculated with realistic 
parameters, as given in Ref.~\cite{SSC:1998}, i.e., $\omega_{p}=2$~eV , 
$t^{+}=0.2$~meV, at $T=20$~K.

\begin{figure}
\includegraphics[width=0.4\textwidth]{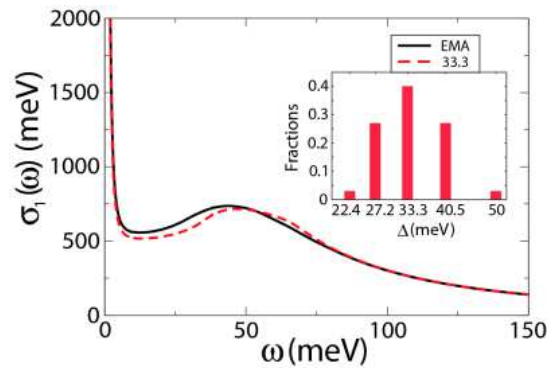}
\caption{\label{fig:dwaveema} (Color online) Solid black line: EMA calculation involving 
constituents described in Fig.~\ref{fig:5d} with volume fractions from Fig.~\ref{fig:gapmaps:c} (also shown as an inset here).  Dashed red curve: an overlay of the $\Delta=33.3$~meV curve of Fig.~\ref{fig:5d} for comparison.}
\end{figure}

The overall characteristics of the resulting EMA in this case (shown in 
Fig.~\ref{fig:dwaveema}) remain what would be expected of a composite of a 
d-wave superconductor with scattering in the clean limit. Indeed, if we
superimpose the conductivity curve for the average gap value, which
dominates the gap distribution seen in STM, we find that there is not much
difference between the two curves and therefore
we conclude from these calculations that even if the inhomogeneity persists 
throughout the bulk, the size of variation in the gap, on scales seen in STM, 
cause very little variation in the overall conductivity of the sample from
that of a single gap system, and 
therefore it is not possible to determine the presence of
inhomogeneities from bulk far-infrared optical measurements.  We note, as well, that 
variation in scattering rates of constituent grains (of up to a factor of 10) 
does not change the qualitative results of an EMA mixture with d-wave symmetry.
We do not expect to have large values of the scattering rate, as this would quench the superconducting state, and there is no evidence from STM or otherwise for normal regions mixed in amongst the regions of inhomogeneous superconductivity at low temperatures.
We conclude that the narrow range of
gap values seen in STM coupled with the rather smooth nature of the
conductivity for a BCS d-wave superconductor makes the observation of
possible gap inhomogeneities uncertain through far-infrared optics.  Adding in realistic inelastic scattering would not change this conclusion as this type of scattering will not introduce new sharp features in the conductivity.  As a final point, most approaches to calculating optical properties of the cuprates assume a homogeneous model.  Our results indicate that it is appropriate to compare results of such calculations for the average gap with experiment, and that ignoring inhomogeneities is a valid first approximation. 

\section{S-wave Energy Gap Inhomogeneities}

\begin{figure}
\subfigure[]{
\includegraphics[width=0.38\textwidth]{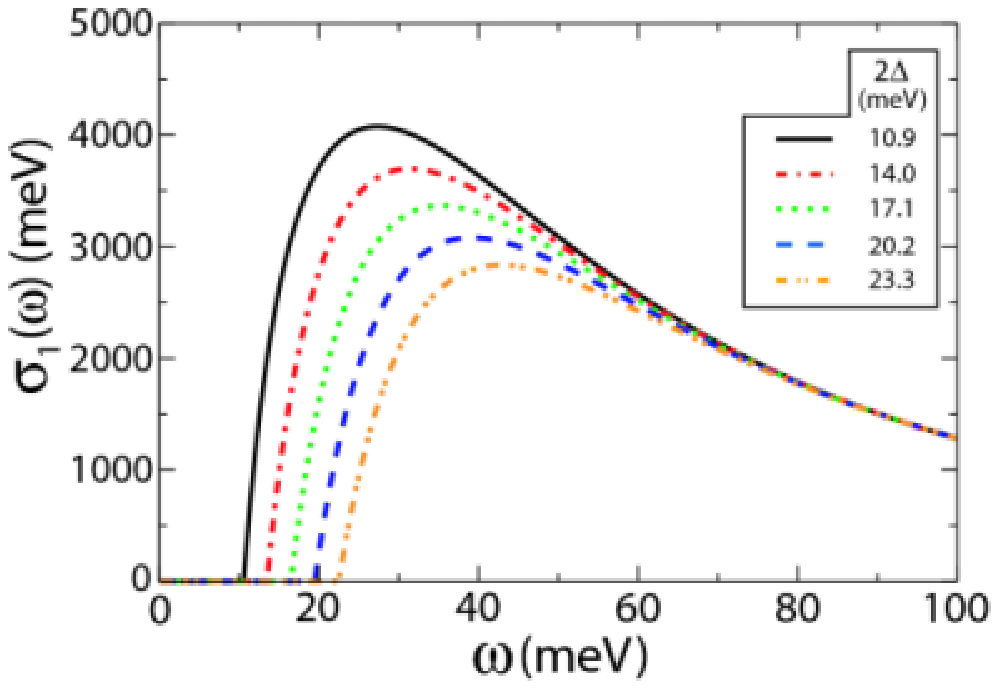}
\label{fig:swavehist:a}
}\\
\subfigure[]{
\includegraphics[width=0.22\textwidth]{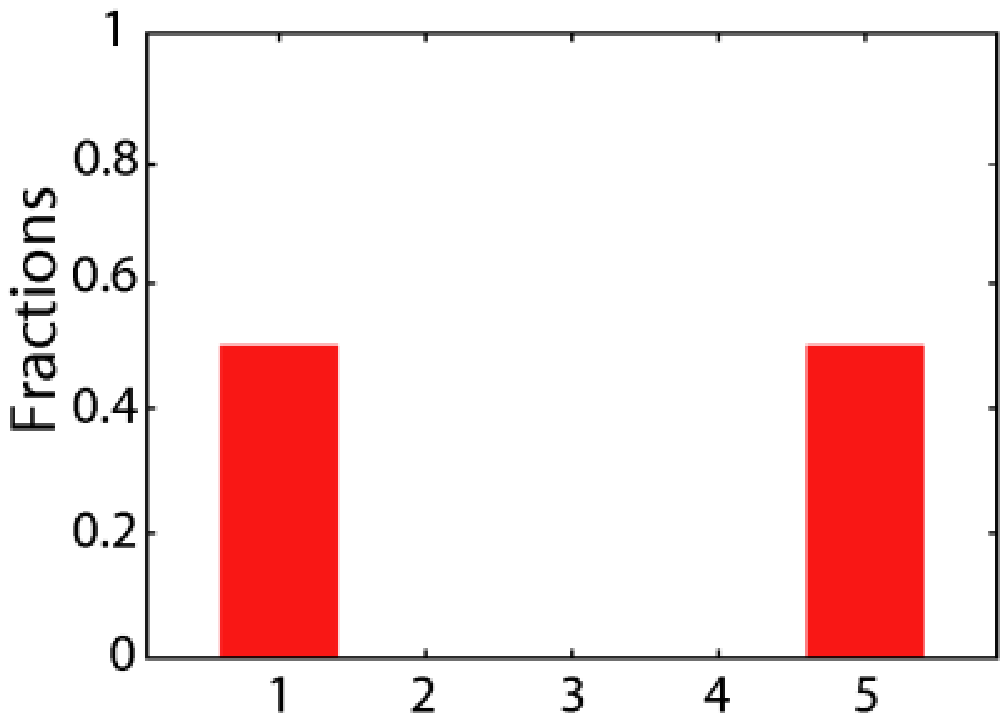}
\label{fig:swavehist:b}
}
\subfigure[]{
\includegraphics[width=0.22\textwidth]{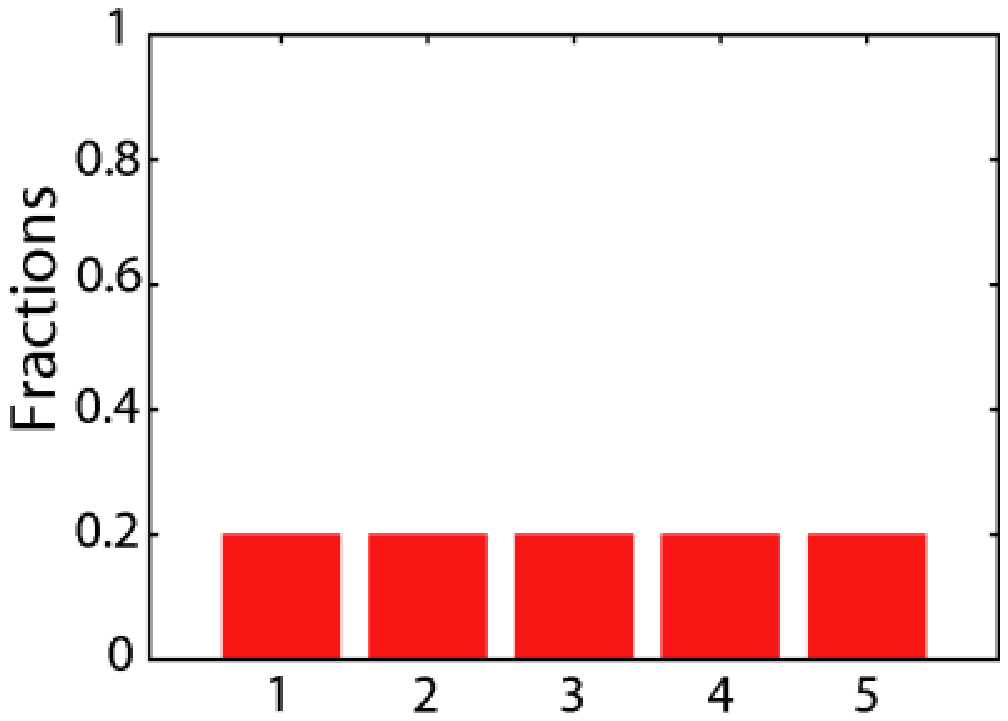}
\label{fig:swavehist:c}
}
\caption{\label{fig:swavehist} (Color online) $\mathbf{(a)}$ 
Real part of the conductivity for different values of the
energy gap
$\Delta$, as discussed in the text.
$\mathbf{(b)}$ Histogram of 2 grain mixture. $\mathbf{(c)}$ Histogram of 5 grain mixture.  
(1$\rightarrow$5 refer to lowest to highest gap values.)}
\end{figure}

With the knowledge of the previous section, we are now motivated to
find possible systems where gap inhomogeneities could have a clear
signature in optics and indeed, it is clear that a system with
a conductivity curve which contains sharp features would be an ideal
candidate. Consequently,
we now examine the effect of a distribution of gaps on a conventional s-wave 
BCS superconductor.  We begin by calculating the zero temperature BCS s-wave 
conductivities for two simple distributions of gap values shown in 
Figs.~\ref{fig:swavehist:b} and \ref{fig:swavehist:c}.  We have assumed a constant elastic scattering
rate $1/\tau=50$ meV, and plasma frequency $\omega_{p}=2$eV for
all grains and this puts the system 
in the regime where $2\Delta<1/\tau$, which is a mildly dirty limit.
We calculate this EMA mixture under the assumption that 
each constituent maintains
 the same overall transition temperature, therefore, variations in $\Delta$ 
value can be seen as analogous to changing the $\frac{2\Delta}{T_{c}}$ ratio 
from 3.53 $\rightarrow$ 7.53 in steps of 1. Curves for the real part of
the conductivity, used for input to the EMA, are shown in Fig.~\ref{fig:swavehist:a}.

\begin{figure}[h]
	\includegraphics[width=0.4\textwidth]{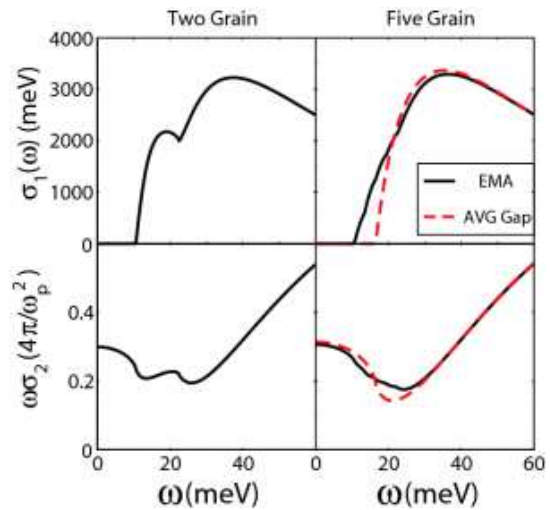}
	\caption{\label{fig:2by2swaves} (Color online)
EMA result for two-grain (left) 
and five-grain (right) composites determined from the histograms in Figs.~\ref{fig:swavehist:b} and \ref{fig:swavehist:c}.
The dashed red curve is the single-gap calculation using the average
gap value of the distribution.}	
\end{figure}

In the two grain composite, Fig.~\ref{fig:2by2swaves} (left column), we 
observe that the separation of gap onsets is sufficient to clearly display a two-gap result in the real part of the conductivity, $\sigma_{1}$ (top frame).  The real part of the conductivity rises sharply out of zero as it does in all of the curves in Fig.~\ref{fig:swavehist:a}.  This is followed by a first peak, after which a second rise begins at twice the value of the largest gap in our two gap distribution.  A second peak follows at higher energy before the normal state Drude form is recovered.  This is fundamentally different from, and should not be 
confused with, a two-band superconductor such as MgB$_{2}$.\cite{nicol:2005}  It should be 
noted that the EMA is distinct in this two-gap system from a fractional 
average of its constituents since both $\sigma_{1m}$ and $\sigma_{2m}$ depend 
on both the real and imaginary parts of the constituents.  In other words, for 
a two component mixture labeled $a$ and $b$, where 
$\sigma_{a}=\sigma_{1a}+i\sigma_{2a}$ and 
$\sigma_{b}=\sigma_{1b}+i\sigma_{2b}$, the EMA results in $\sigma_{1m}=\sigma_{1m}(\sigma_{1a},\sigma_{2a},\sigma_{1b},\sigma_{2b})$ and $\sigma_{2m}=\sigma_{2m}(\sigma_{1a},\sigma_{2a},\sigma_{1b},\sigma_{2b})$.  This additional dependence of the real part of the
conductivity on the imaginary parts of 
its constituents in general leads to an increased low frequency contribution 
to $\sigma_{1m}$ from the imaginary conductivities (physically seen as 
scattering from constituents).  The imaginary part multiplied by $\omega$, namely $\omega\sigma_{2}(\omega)$, is shown in the lower left panel.  As the frequency goes to zero, we find the inverse square of the London penetration depth for the composite system.  At finite frequency, there are structures in $\omega\sigma_{2}(\omega)$ corresponding to the frequencies of the onset of absorption of each of the two grains included in the model.  These structures would be easily identifiable.

When we consider a less-separated five-grain distribution [seen in 
Fig.~\ref{fig:swavehist:c}, with results in Fig.~\ref{fig:2by2swaves} 
(right column)], we immediately recognize the loss of clear onset points for 
each gap value and instead see a broadened rise in $\sigma_{1}(\omega)$ (Fig.~\ref{fig:2by2swaves}, top right frame), which quickly 
rejoins an average gap fit at frequencies just above the 
highest $2\Delta$ point. Shown here is the EMA calculation as the solid
black curve with the dashed red curve being the calculation for a single
gap system using the average gap of the distribution as the input gap.  It is clear that in this more realistic case the signature of inhomogeneities is not as readily identifiable as for the two gap case.  This is also true for the imaginary part, $\omega\sigma_{2}(\omega)$, shown in the lower frame; now a single minimum is seen similar to the single gap case, but somewhat more broadened.

%Replaced Text with jules'

%In order to see this discrepancy from a single gap fit more clearly, we may 
%wish to consider the optical self-energies defined as permutations of 
%$\sigma_{1}$ and $\sigma_{2}$ as:\cite{carbotte:2005,gotze:1972}
%\begin{equation}
%\frac{1}{\tau_{op}}=\frac{\omega_{p}^{2}}{4\pi}
%\frac{\sigma_{1}(\omega)}{\sigma_{1}^{2}(\omega)+\sigma_{2}^{2}(\omega)}
%\label{eqn:tauopt}
%\end{equation}
%and
%\begin{equation}
%1+\lambda_{op}=\frac{\omega_{p}^{2}}{4\pi\omega}
%\frac{\sigma_{2}(\omega)}{\sigma_{1}^{2}(\omega)+\sigma_{2}^{2}(\omega)}.
%\label{eqn:lamopt}
%\end{equation}

%When the results of Fig.~\ref{fig:2by2swaves} are permuted as in 
%Eqs.~(\ref{eqn:tauopt}) and (\ref{eqn:lamopt}),
% we see clearly that the gap onset 
%in plots of $1/\tau_{op}$ is much slower than a single gap curve, while in the 
%plot of $1+\lambda_{op}$, we see the clear location of $2\Delta_{eff}$.  In 
%general, the mass renormalization peaks at a lower frequency and has a lower 
%magnitude than a single gap system of comparible volume averaged gap.  We 
%conclude from this that in s-wave mixtures, distributions of gaps should 
%display themselves distinctly from single gap systems in terms of a broadened 
%onset in $\sigma_{1}$ or $1/\tau_{op}$ and also a decreased effective mass, 
%$1+\lambda_{op}$, in composite systems.

\begin{figure}[h,t,b]
	\includegraphics[width=0.4\textwidth]{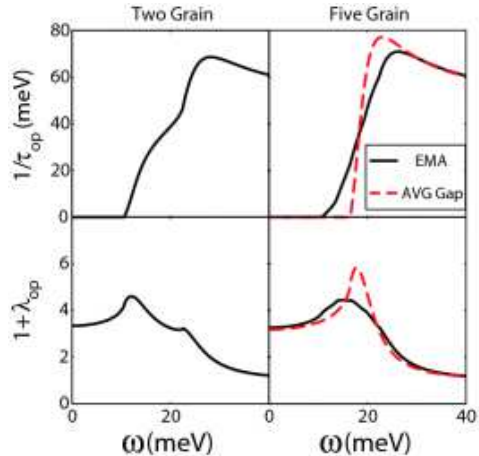}
	\caption{\label{fig:5taulam} (Color online)
EMA result for the theoretical self-energy of the two grain (left) and five
grain (right) composite determined by histograms in Figs.~\ref{fig:swavehist:b} and \ref{fig:swavehist:c}, respectively. Top row shows the scattering rate, $1/\tau^{op}$, and the bottom row shows the mass renormalization, $1+\lambda_{op}$.}	
\end{figure}

The real and imaginary parts of the conductivity are not the only optical quantities that enter a more modern discussion of the subject.  In fact, it has been found to be very useful to introduce an optical self-energy $\Sigma^{op}(\omega)$ defined in terms of the generalized Drude formula for the optical conductivity, $\sigma(\omega)$, which is written as:\cite{carbotte:2005,gotze:1972,mori:2008}
\begin{equation}
\sigma(\omega)=\imath\frac{\omega^{2}_{p}}{4\pi}\frac{1}{\omega-\Sigma^{op}(\omega)}.
\label{eqn:drude}
\end{equation}
Here $\omega_{p}$ is the plasma frequency and $\Sigma^{op}(\omega)$ has both a real and imaginary part related, respectively, to a frequency dependent optical effective mass, $m^{*}_{op}(\omega)/m$, (where $m$ is the bare electron mass) and scattering rate, $1/\tau^{op}(\omega)$, given by:
\begin{equation}
\omega[m^{*}_{op}(\omega)/m-1]=-2\Sigma^{op}_{1}(\omega)
\end{equation}
and
\begin{equation}
1/\tau^{op}(\omega)=-2\Sigma^{op}_{2}(\omega).
\end{equation}
An optical mass renormalization parameter, $\lambda^{op}(\omega)$, is then defined by $m^{*}_{op}(\omega)/m=1+\lambda^{op}(\omega)$.  These quantities are analogous, but different from, the quasiparticle mass renormalization and scattering rate which follow from the quasiparticle self-energy $\Sigma^{qp}(\omega)$.  Instead, $\Sigma^{op}(\omega)$ is a two-particle quantity~\cite{hwang:nicol:2007}.  It is related to the Kubo formula for the current-current correlation function which determines $\sigma(\omega)$.  In general, as stated, the optical and quasiparticle mass renormalization and scattering rate are not the same, although in some limits they can be.  As an example, the zero frequency mass renormalization $\lambda^{op}(\omega=0)$ and $\lambda^{qp}$ are equal.
In the end, of course, $\Sigma^{op}(\omega)$ contains exactly the same information as does $\sigma(\omega)$.  In fact, one can solve for $1/\tau^{op}$ and $\lambda_{op}$ in terms of $\sigma_{1}$ and $\sigma_{2}$ to find:\cite{carbotte:2005,gotze:1972}
\begin{equation}
\frac{1}{\tau_{op}}=\frac{\omega_{p}^{2}}{4\pi}
\frac{\sigma_{1}(\omega)}{\sigma_{1}^{2}(\omega)+\sigma_{2}^{2}(\omega)}
\label{eqn:tauopt}
\end{equation}
and
\begin{equation}
1+\lambda_{op}=\frac{\omega_{p}^{2}}{4\pi\omega}
\frac{\sigma_{2}(\omega)}{\sigma_{1}^{2}(\omega)+\sigma_{2}^{2}(\omega)}.
\label{eqn:lamopt}
\end{equation}
This does not mean, however, that these quantities have no particular value of their own.  It is well documented that they can speak more directly to certain questions than can $\sigma$ itself.  For example, $1/\tau^{op}(\omega)$ provides information on absorption and, in an s-wave superconductor, at zero temperature it is zero for photon energies, $\omega$, less than twice the gap.  A coherence peak is seen above this energy and at high $\omega$ we recover a measure of the normal state elastic scattering.  Further, it has recently been found that in the underdoped cuprates, the real part of $\Sigma^{op}(\omega)$ possesses a ``hat like'' structure which extends in energy over a range of about twice the pseudogap energy.  This characteristic structure, which is superimposed on a large smooth incoherent background is nevertheless unmistakable and provides a direct image of pseuodogap formation \cite{hwang:2008} as it enters the in-plane optical data.  By contrast, the effect of the opening of a pseudogap in the real and imaginary part of the conductivity is much more subtle and is spread over a much larger energy range.  Thus, the optical quantities defined in Eq.~(\ref{eqn:tauopt}) and (\ref{eqn:lamopt}) have proved useful and here we will consider how they are changed in the case of a distribution of regions of distinct gap values.  In the top row of Fig.~\ref{fig:5taulam} we show results for $1/\tau^{op}(\omega)$ (solid black curve) in meV as a function of $\omega$ (also in meV), for the two- and five-gap distributions of Figs.~\ref{fig:swavehist:b} and \ref{fig:swavehist:c}.  The onset of scattering starts at the lowest value of twice the gap which is roughly 11 meV in our model.  Also shown for comparison (dashed red curve) are results for the case of a homogeneous superconductor with the same average gap value.  We see that, by comparison, the rise in $1/\tau^{op}(\omega)$ in the inhomogeneous case, is much less steep and proceeds more gradually, being spread out over an energy scale which corresponds to the spread in gap values in our model distribution.  The peak in the red dashed curve has its origin in the well known coherence peak of the BCS s-wave electronic density of states.  In the normal state, $1/\tau^{op}(\omega)$ would be independent of $\omega$ and equal to the input constant impurity scattering rate of 50~meV in this example.  This is the value to which $1/\tau^{op}(\omega)$ (superconducting) tends towards for $\omega$ greater than a few times the gap.  In the superconducting state, however, no scattering is possible below twice the gap, at which energy it rises sharply and overshoots its normal state value, not because the intrinsic scattering potential has increased but rather because there are more final states available for scattering.  For the inhomogeneous case this feature is simply smeared out somewhat because of the distribution of different superconducting regions which are sampled.

In the lower row of Fig.~\ref{fig:5taulam} we show our results for the optical mass enhancement parameter $[1+\lambda_{op}(\omega)]$ in the two- and five-gap cases.  The solid black curve is for the inhomogeneous case and the red dashed curve is the homogeneous case with an averaged gap value for comparison.  We note the inhomogeneities smear out the peak in $\lambda_{op}(\omega)$ around $2\Delta$.  However, in all other aspects the curves are similar.  We emphasize two points.  First, for $\omega$ large compared with the gap, the mass renormalization factor $\lambda_{op}(\omega)$ tends towards zero, which agrees with its normal state value.  At $\omega=0$ however, $\lambda_{op}$ is seen to be quite large ($\sim$2.5) in both the homogeneous and inhomogeneous cases.  In this limit, the optical effective mass has a very definite meaning.  It is the value of the electron mass that one is to give the electrons if one wishes to use the classical London formula for the penetration depth $\lambda_{L}$ at T=0, namely $\lambda^{-2}_{L}=4\pi ne^{2}/m^{*}c^{2}$ where n is the free electron density, e the electron charge and c the velocity of light.  In the clean limit, i.e., no elastic scattering, $m^{*}$ would be the bare electron mass.  Including impurities in BCS theory reduces the London penetration depth and this can be expressed by changing $m$ to $m^{*}(\omega=0)$ in the conventional expression.  Our results show that this result is only slightly changed in the inhomogeneous case, i.e. $m^{*}(\omega=0)$ is only very slightly larger.  The main effect of inhomogeneities manifests itself at finite frequency in the region of twice the gap.  We see that, as compared with the dashed red curve, the solid black curve shows a much broader peak which is a direct consequence of spatial gap variations.  The peak in $\lambda^{op}(\omega)$ around twice the gap is directly tied to the rapid rise in $1/\tau^{op}(\omega)$ at this same frequency, as these are Kramers-Kronig (KK) related, i.e., a step in the scattering rate at $\omega=2\Delta$ translates into a logarithmic singularity at $2\Delta$ in its KK transform.  We note in passing that a recent optical study of MoGe (Ref.~\cite{tashino:2008}) has revealed such a smearing of coherence peaks in the thinnest films studied, which could be attributed to inhomogeneities, however, other evidence suggests otherwise.

\section{Summary and Conclusions}

Optical absorption is a probe of bulk properties.  Using effective medium 
theory, we have calculated the optical conductivity of a superconductor 
consisting of a random array of nanoscale regions having different magnitudes 
of superconducting gap.  The superconductivity in each grain is described 
within a BCS model, and both s- and d-wave symmetry is considered.  For 
simplicity, models consisting of two and five distinct gap values are 
examined in detail.  For an s-wave superconductor, the real part of the 
optical conductivity is zero for frequencies below twice the gap, with the 
missing optical spectral weight transferred to the condensate.  For d-wave, 
the region below $2\Delta$ is also depleted, although some absorption remains 
at all frequencies and there is no sharp threshold in $\sigma_{1}(\omega)$ 
vs $\omega$.  In this latter circumstance the conductivity of the composite system 
does not exhibit qualitatively different behaviour as compared with the single 
grain which would represent a clear signature of the bulk inhomogeneities.  
In fact, there is no quantitative differences between $\sigma(\omega)$ of the 
composite and the individual grains, and the electromagnetic response 
appears close to that for the average gap.  By contrast, for an s-wave gap, 
the sharp absorption edge in $\sigma_1(\omega)$ at twice the gap which 
is characteristic of a uniform BCS superconductor becomes less steep reflecting
the onset of a distribution of gaps rather than of a single one.  For a 
distribution involving only two well separated gap values, each is seen 
distinctly in both the real and imaginary parts of the conductivity as well 
as in the real and imaginary parts of the corresponding optical self energy.  
This latter property follows directly from a knowledge of the conductivity and 
has been found useful in past discussions of, for example, boson structures 
in optical properties.  For the case of a distribution of many gap values, 
the presence of each individual gap is less evident with $\sigma_{1}(\omega)$ 
showing rather gradual onset over a frequency range representative of the 
variation in gap values. In conclusion, signatures of nanoscale spatial 
inhomogeneities in the far-infrared optical conductivity 
are much more readily recognized for composites 
consisting of s-wave gap symmetry components than they are for those 
with d-wave gap symmetry.  Indeed, the fact that the d-wave case appears to be unaffected by the existence of inhomogeneities in spite of their apparent presence in STM, gives support for the continued use of single gap models for calculation and analysis of optical conductivity in the high-$T_{c}$ cuprates.

%\vspace{10pt}
%\newpage
\begin{acknowledgments}
We thank Ewald Schachinger and Ricardo Lobo for providing
some of the computer programs used in this work. We also thank Kyle McElroy
for providing us with the gap map of Fig.\ref{fig:gapmaps:a} and
Clare Armstrong for assistance with the schematic diagram of Fig. \ref{fig:gapmaps:b}.  This work has been supported by the Natural Sciences and
Engineering Council of Canada (NSERC) and the Canadian Institute
for Advanced Research (CIFAR).
\end{acknowledgments}
%\appendix
%\section{Appendixes}
%\newpage %Just because of unusual number of tables stacked at end
\bibliographystyle{apsrev}
\bibliography{bib2}% Produces the bibliography via BibTeX.

\end{document}